# Transfer Printing Approach to All-Carbon Nanoelectronics

V. K. Sangwan*[1,2], A. Southard[1], T. L. Moore[1,2], V. W. Ballarotto[2], D. R. Hines[2], M. S. Fuhrer[1], E. D. Williams[1,2]

1. Center for Nanophysics and Advanced Materials, University of Maryland, College Park, MD 20742
2. Laboratory for Physical Sciences, College Park, MD 20740

**Abstract**

Transfer printing methods are used to pattern and assemble monolithic carbon nanotube (CNT) thin-film transistors on large-area transparent, flexible substrates. Airbrushed CNT thin-films with sheet resistance 1kΩ square$^{-1}$ at 80% transparency were used as electrodes, and high quality chemical vapor deposition (CVD)-grown CNT networks were used as the semiconductor component. Transfer printing was used to pre-pattern and assemble thin film transistors on polyethylene terephthalate (PET) substrates which incorporated $Al_2O_3$/poly-methylmethacrylate (PMMA) dielectric bi-layer. CNT-based ambipolar devices exhibit field-effect mobility in range 1 – 33 cm$^2$/Vs and on/off ratio ~$10^3$, comparable to the control devices fabricated using Au as the electrode material.

* Current address: Materials Science and Engineering, Northwestern University, IL 60208

## I. INTRODUCTION

Carbon nanotube (CNT) thin-films have been successfully incorporated as both high-quality semiconductor [1-6] layers and electrodes [6-10] in large area flexible, transparent electronics. Semiconducting CNT thin-films are transparent, flexible, environmentally stable, and possess higher field-effect mobility than organic semiconductors (e.g. pentacene and P3HT) [5, 11]. However, use of metal electrodes (e.g. Au, Ti and Pd) make CNT thin-film transistors (TFTs) only partially transparent and sub-optimally flexible. Conventional transparent, conducting films (e.g. indium tin oxide (ITO)) have been used as electrode materials in organic TFTs [12] and CNT TFTs [13], but expensive growth techniques and inherent brittleness make these materials incompatible with flexible electronics. Alternatively, solution-processed CNT thin-films have been optimized to achieve sheet resistance and transparency that are comparable to ITO, as well as being more flexible than ITO [7, 8, 14, 15]. Thus, CNT thin-films have been independently utilized as high quality active components and as electrodes. However, there have been limited efforts [6] to combine these two functionalities of CNT thin-films together to achieve carbon-based, all-transparent flexible electronics.

In this paper, we report on a novel transfer printing approach to assemble all-CNT devices using CVD-grown CNT thin-films as the active semiconducting layer and solution-processed CNT thin-films as the electrodes. A transfer printing method [16-19] was used to pattern the semiconducting CNT thin-film without exposure to any processing chemicals while an airbrushing method was used to produce the CNT thin-film electrodes. The airbrushing method allowed for rapid production of large area thin-films on a variety of substrates at room-temperature [20]. The sheet resistance of the CNT electrodes was ~ $1 k\Omega\, square^{-1}$ at 80 % transparency and they exhibited superior flexibility as compared to conventional transparent conducting films such as doped $In_2O_3$. Gate leakage in printed CNT TFT devices was avoided by engineering an organic/inorganic hybrid dielectric. An $Al_2O_3$/poly-methylmethacrylate (PMMA) dielectric bi-layer was used to achieve a minimal gate-leakage near the resolution of the measurement set up (>100 pA). CNT-based devices on a polyethylene terephthalate (PET) substrate exhibited field-effect mobilities in the range 1 - 33 $cm^2/Vs$ and on/off ratios up to $10^4$. In contrast to *p*-type control devices, these CNT-based devices showed ambipolar behavior that could be useful in complementary circuits.

## II. FABRICATION

### A) Semiconducting Carbon Nanotube Thin Films

Semiconducting CNT thin-films were grown on thermally oxidized, Si transfer substrates by CVD using iron nanoparticles as a catalyst [19]. A transfer printing approach was used to pattern the semiconducting CNT thin-films, as described in Ref [19]. Briefly, the growth substrate containing a CNT thin-film was pressed against a PET stamp at 400 psi, 140 ºC for 3 min. CNTs in direct contact with the raised parts of the PET stamp were transfer printed while CNTs not in contact with the stamp (under recessed areas of the PET stamp) remained unaffected on the growth substrate. The resulting patterned CNT thin-film consisted of isolated areas 200 μm x 300 μm in size with edge roughness on the order of ± 5 μm.

To facilitate the transfer printing of the patterned CNT thin-films, the surface energy of uncovered areas of the $SiO_2$/Si transfer substrates was reduced using a self-assembled monolayer (SAM) treatment method, as described in Ref [19]. Briefly, areas of the $SiO_2$/Si substrate uncovered by the patterning of the CNT thin-film were treated with a (tridecafluoro-1,1,2,2-tetrahydrooctyl) trichlorosilane SAM. This SAM, acting as a release layer, decreases the adhesion between the transfer and device substrates [16]. The release layer, however, should only cover bare regions of the transfer substrate and not the patterned areas of the CNT thin-film. This was achieved by coating a polydimethylsiloxane (PDMS) stamps with the release layer molecules and using micro-contact printing. The release layer-coated PDMS stamp was aligned and placed over the patterned CNT thin-film such that only raised areas of the PDMS stamp were in contact with the bare substrate. The PDMS stamp was gently removed after 15 min.

**B) Carbon Nanotube Thin Film Electrodes**

For the fabrication of electrodes, CNT powder (P3, Carbon Solutions, Inc.) was used as-purchased. These CNTs are functionalized with 4-6 atomic% carboxylic acid at 80-90% carbonaceous purity and were dispersed at a concentration of 1 mg/mL in a solution of 1% by wt. sodium dodecyl sulfate (SDS) in de-ionized (DI) water [9]. The CNT solution was first bath sonicated for 90 min and then centrifuged at 12,000 rpm for 40 min to remove impurities. The resulting well-dispersed, purified CNT solution was airbrushed (Aztek A470 airbrush kit, 40 psi) onto a desired substrate, in this case a Au coated Si wafer (at 165 $^{o}$C) and then soaked in DI water for 30 min to remove the surfactant. It was observed that airbrushing a dilute CNT solution using multiple short pulses (approximately 0.5 sec each) produced a more uniform thin-film compared to that produced by a continuous spray. Transparency versus sheet resistance measurements of these CNT thin-films were previously reported in Ref [9]. The electrodes reported here are approximately 30 nm-thick with a sheet resistance of 1 k$\Omega$ square$^{-1}$ at 80% transparency (at wavelength of 550 nm).

The transfer printing method from Section IIA could not be employed to pattern these airbrushed CNT thin-films because they transferred only partially onto a PET substrate, possibly due to a stronger CNT/Si adhesion strength compared to cohesion strength of 30 nm thick multilayered CNT thin-films. Thus, airbrushed CNT thin-films were patterned into electrodes using photolithography and $O_2$ plasma reactive ion etching (RIE) [19]. No significant change in sheet resistance was observed after patterning. Since, airbrushed CNTs transfer partially onto device substrates a carrier film was used to ensure complete transfer of CNT thin-film electrodes as follows. The CNTs were airbrushed and patterned onto a 50 nm-thick Au carrier film deposited by thermal evaporation onto a Si transfer substrate (Fig. 1(a)). As compared to the CNT thin-film, the Au carrier film has a weaker adhesion to the Si substrate, but yet a strong enough cohesion to allow complete and intact transfer of the CNT thin-film/Au bi-layer to a PET device substrate [16]. Other approaches such as airbrushing CNTs on a release layer-treated Si were investigated, but the resulting thin-films were too fragile to survive soaking in DI water. Using Au carrier films, we have also achieved proof-of-concept transfer printing of airbrushed CNTs to other polymer dielectric materials such as poly(4-methylstyrene), poly(alpha-methylstyrene), polystyrene, poly-4-hydroxystyrene, polyvinyl alcohol, polycarbonate, polyimide and polyvinyl nitrile.

**C) Assembling All-Carbon Nanotube Devices**

For all-CNT bottom gate TFTs, different device components were assembled in the following order: airbrushed and patterned CNT thin-film gate electrodes, gate dielectric layer, airbrushed and patterned CNT thin-film source-drain electrodes, and finally, CVD-grown and patterned CNT thin-films as the active layer. The CNT gate electrodes/Au bi-layer (discussed above in Section IIB) was transfer printed from the Si transfer substrate to a PET device substrate at 400 psi, 120 $^{o}$C for 3 min (Fig. 1(b)). The Au film was then removed from the PET substrate by soaking in a solution of Au etchant (GE-1848, Transene Company, Inc.) for 1 minute. Neither optical images nor electrical resistance measurements showed any discernible effect of the wet chemistry Au etch on the CNT gate electrode or the PET substrate.

The gate dielectric layer was prepared by depositing a 50 nm thick $Al_2O_3$ layer onto the gate electrodes by e-beam evaporation followed by spin-coating an 800 nm-thick PMMA layer (Fig. 1(c)). Without the $Al_2O_3$ layer, gate leakage was observed even with a 2 μm thick stand-alone PMMA layer. The $Al_2O_3$ is used as a diffusion barrier for CNTs that appear to diffuse through the bulk of the polymer dielectric layer at elevated temperatures associated with the transfer printing process.

The CNT source and drain electrodes/Au bi-layer was printed onto the PET substrate containing the CNT gate electrode and PMMA/$SiO_2$ dielectric layer. The source/drain electrodes were printed onto the top surface of the PMMA layer at 400 psi, 120 $^{o}$C for 3 min (Fig. 1(d)) followed by etching of the Au film by Au etchant. In the final step, a CVD-grown and patterned CNT thin-film (see Section IIA) was transfer printed from a release layer treated $SiO_2$/Si transfer substrate to the electrode sub-assembly at 500 psi, 170 $^{o}$C for 3 min (Fig. 1(e)).

To demonstrate the transfer printing technique for large area electronics on transparent, flexible substrates, 90 CNT electrode sub-assemblies were fabricated on a one square-inch PET substrate (Fig. 2(a)). Bottom gate CNT-based devices were fabricated on 5 mm x 5 mm PET substrates with channel width (*W*) of 100 μm and channel length (*L*) varying from 10 μm to 100 μm in steps of 10 μm. The SEM image in Fig. 2(b) shows a semiconducting CNT thin-film connected to CNT source-drain electrodes across a 20 μm long channel. Individual CNTs in the denser source-drain electrodes cannot be identified in the SEM image due to charging of the plastic substrate.

### D) Assembling Control Devices

CNT thin-film reference devices were fabricated using the transfer printing method described in Ref. [19]. Briefly, 100 nm-thick Au gate electrodes were first transfer printed onto a PET device substrate. A 50 nm thick $Al_2O_3$ layer was then deposited over the gate electrodes and covered with an 800 nm thick spin-coated PMMA layer. 30 nm thick Au source-drain electrodes were then printed onto the device substrate. Finally, patterned CVD-grown CNT thin-films were printed onto the source-drain electrodes. All the transfer printing steps were carried at 500 psi, 170 $^{o}$C for 3 min.

## III. RESULTS

### A. Flexibility of Carbon Nanotube Thin Films

First, we report on flexibility of airbrushed CNT thin-films. PET stripes coated with airbrushed CNT thin-films were wrapped around cylinders of varying diameters to induce tensile strain. The variation in sheet resistance ($R_f/R_i$; where $R_i$ and $R_f$ are sheet resistance before and after bending) as a function of radius of curvature ($r$) is plotted in Fig. 3 for tensile stress (CNT electrodes on the outer side of bent PET; black circles) and compressive stress (CNT electrodes on the inner side of bent PET; red squares). In both cases, sheet resistance starts changing at $r$ = 10 mm, but changes only by 7% when the substrate is bent to $r$ = 2 mm ($R_{i,CNT}$ = 500 Ωsquare$^{-1}$) (see inset of Fig. 3). Multiple bending tests were also performed and sheet resistance changed only by 12% after bending to $r$ = 2 mm for 35 times. In addition, two

CNT-based device were measured before and after the bending and no significant change in transport properties was observed up to $r = 6$ mm. Bending the devices further ($r < 6$ mm) resulted in gate leakage. Thus, the PMMA/$Al_2O_3$ bi-layer acts as an effective diffusion barrier of CNTs under significant strain. Flexibility of airbrushed CNT thin-films was compared with that of commercially available $In_2O_3$/Au/Ag film (Delta Technologies Limited). A PET film sputtered with $In_2O_3$/Au/Ag showed three orders of magnitude increase in sheet resistance ($R_{i,In2O3} = 10$ Ωsquare$^{-1}$) for $r = 3$ mm, Fig. 3. Thus, CNT thin-films are observed to have a superior flexibility than doped $In_2O_3$ films.

**C. Transport Measurement**

Transport measurements of all-CNT devices were made in ambient conditions using a Cascade Microtech probe station. 100 nm-thick Au pads were deposited on the CNT electrodes 500 μm away from the device channel to make better contact with the probe tips. The resistance of the 500 μm-long CNT source and drain electrodes was subtracted from the total device resistance to obtain true *I-V* characteristics. Some devices show a poor on/off ratio due to the presence of metallic CNTs in the random network. We employed an electrical breakdown technique [21] to eliminate metallic paths in the CNT thin-films. To achieve that, drain bias was increased in controlled fashion while keeping the gate bias at 30 V. Fig. 4(a) shows the resulting intermediate transfer characteristics after successive burnout of metallic CNTs in a device with $L = 60$ μm and $W = 100$ μm. The on/off current ratio (at $V_d = -1$ V) of the device has increased from 10 to 6000, whereas field-effect mobility (in linear regime, $-1$ V $< V_d < 1$ V) of the device decreased from 10.3 cm$^2$/Vs to 4.80 cm$^2$/Vs. Fig. 4(b) and (c) show output and transfer characteristics of the device after electrical breakdown. Thus, all-CNT bottom gate devices show ambipolar behavior with hole conductivity up to an order of magnitude greater than electron conductivity. In contrast, previously reported control bottom gate devices using Au electrodes had shown *p*-type behavior [19]. After burning metallic paths, on/off ratio of some of the devices was improved by up to 2 – 3 orders of magnitude and no *L* dependence was observed in final on/off ratio. In Fig. 4(d), on/off ratio is plotted as a function of field-effect mobility for CNT-based devices and control devices. The on/off ratio of CNT-based devices is up to two orders of magnitude higher than that previously reported for similar CNT-based devices [6]. The field-effect mobility of CNT-based devices varies between 1 – 33 $cm^2/Vs$ and that of control devices is between 3 – 20 $cm^2/Vs$. Overall, the device performance of CNT-based devices on plastic is comparable to the control CNT thin-film devices reported here and elsewhere [4, 19, 20]

## IV. DISCUSSION

A novel transfer printing approach via a carrier layer was developed to assemble all-CNT devices by incorporating solution-processed CNT electrodes with CVD-grown CNT active layer. This method does not depend on adhesive properties of the printable layer, but relies only on differential adhesion between a carrier layer and the substrates. Here, we used Au as a carrier layer and achieved 100% transfer of CNTs onto 10 different plastic substrates. For electrodes, as-purchased CNT solutions produced thin-films of sheet resistance of 1 kΩ/square at 80% transparency and show a much superior flexibility than doped $In_2O_3$. Though, there is room for improvement in sheet resistant the proposed transfer printing is compatible with high quality solution-processed CNTs demonstrated elsewhere [22, 23].

Transfer printing CNT electrodes directly onto PMMA dielectric layer results in gate leakage (up to 10 μA) due to diffusion of CNTs in PMMA. The gate leakage was observed through 30 nm thick airbrushed CNT electrodes, not through the sub-monolayer of CVD-grown CNTs. The gate leakage was significantly reduced (~ 100 pA, comparable to sensitivity of measurement set-up) by incorporating a diffusion barrier ($Al_2O_3$) in the gate dielectric. Thus, our $Al_2O_3$/PMMA dielectric bi-layer results in significantly lowered gate leakage compared to a $Al_2O_3$/SU-8 dielectric bi-layer reported elsewhere [6]. Though a thin-film of $Al_2O_3$ could be sufficient to achieve reduced gate leakage and higher gate capacitance, the role of the PMMA layer is necessary to achieve the necessary differential adhesion for transfer printing.

Device performance of CNT-based bottom gate devices (field-effect mobility between 1 – 33 $cm^2/Vs$ with typical on/off ratio ~ $10^3$) is comparable to Au-contact bottom gate CNT devices reported elsewhere [19]. CNT-based bottom gate devices showed ambipolar behavior in contrast to *p*-type response of Au-contact bottom gate CNT devices [19]. CNTs have been suggested to be *p*-doped by trap charges in dielectric layer [24, 25] and dopants in ambient environment [26]. Ambipolar behavior of CNT-based devices suggests the possibility of lower doping from the PET substrate compared to $SiO_2$, and also indicates that CNT-CNT contacts are effective in partially retaining the intrinsic behavior of CNTs in ambient conditions. The lower on/off ratio in ambipolar devices compared to unipolar CNT devices could be due to overlap of sub-threshold regimes of electron and hole conducting states [1, 5]. However, comparable device performance and ambipolar behavior in CNT-based devices suggest at least comparable, and possibly improved, contact resistance in CNT-based devices than Au-contacted control devices.

## V. CONCLUSION

In conclusion, we have demonstrated transfer printing as a viable method to achieve all-carbon electronics. Solution-processed CNTs as electrodes and pristine CVD-grown CNTs as active layer were incorporated in bottom-gate CNT-based TFTs on large area PET substrate. A carrier layer-assisted transfer printing approach achieved 100% transfer of CNTs to varieties of substrates, and has a potential of generalization to other nanomaterials. Resulting CNT-based devices show similar performance to Au-contacted control devices with field-effect mobility in range 1 – 33 $cm^3/Vs$ and typical on/off ratio $10^3$. Ambipolarity of CNT-based devices (compared to *p*-type control devices) suggests that these CNTs remain relatively undoped even in ambient conditions.

**Acknowledgements:**

This work has been supported by the Laboratory for Physical Sciences, and by use of the UMD-MRSEC Shared Equipment Facilities under grant # DMR 05-20471. Infrastructure support is also provided by the UMD NanoCenter and Center for Nanophysics and Advanced Materials.

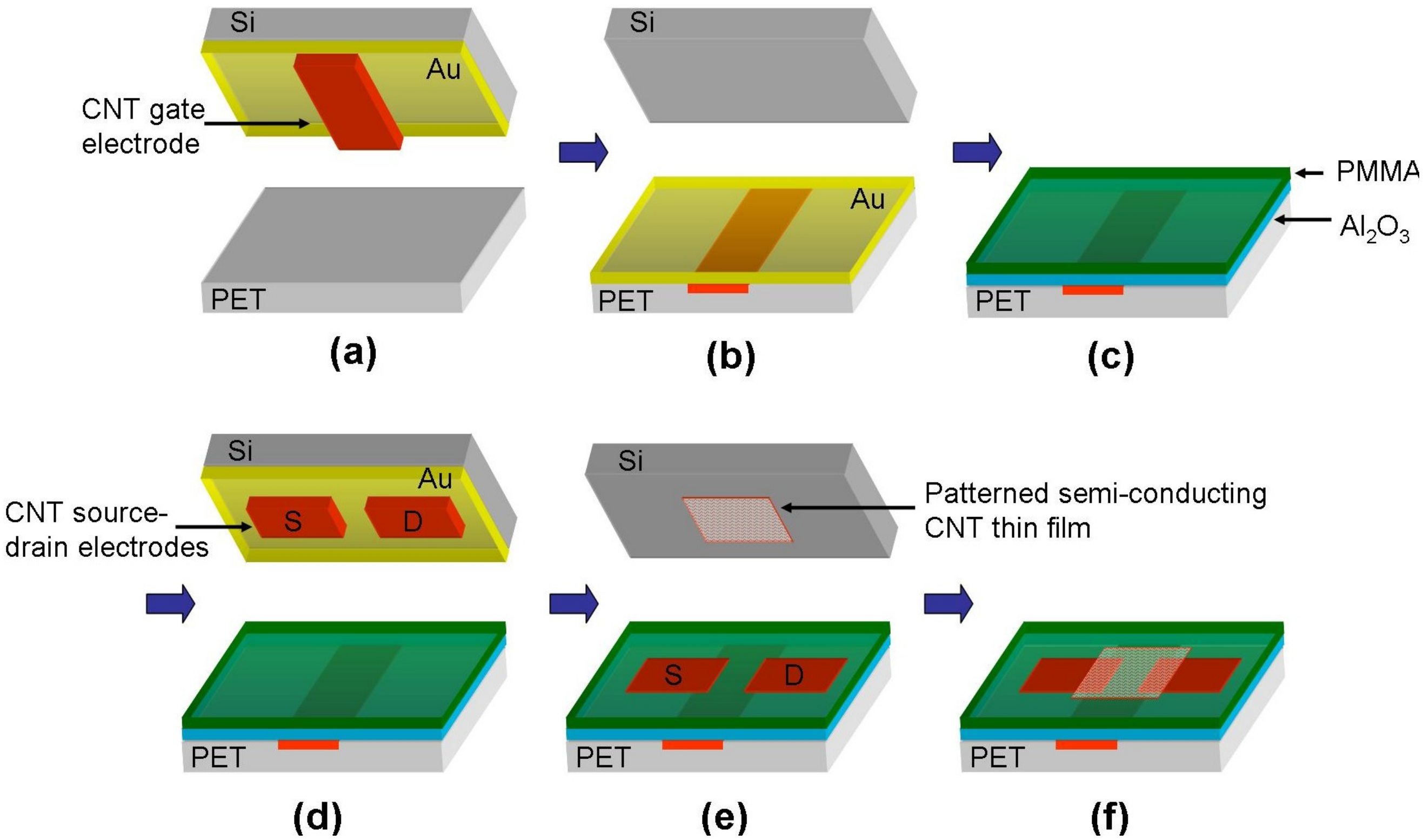

Figure 1. A schematic of different transfer printing steps used to fabricate CNT-based bottom gate devices. a) A 50 nm thick Au carrier film is used to transfer print patterned CNT thin-film electrodes onto a PET device substrate. b) The Au carrier film is chemically etched away. c) A 50 nm thick layer of $Al_2O_3$ was evaporated on gate electrodes and then spin-coated with an 800 nm thick PMMA layer. d) CNT source-drain electrodes were transfer printed onto the PMMA layer using another Au carrier film. e) and f) In the last step, the active component of the device, patterned pristine CNT thin-film was transfer printed onto the CNT electrode sub-assembly.

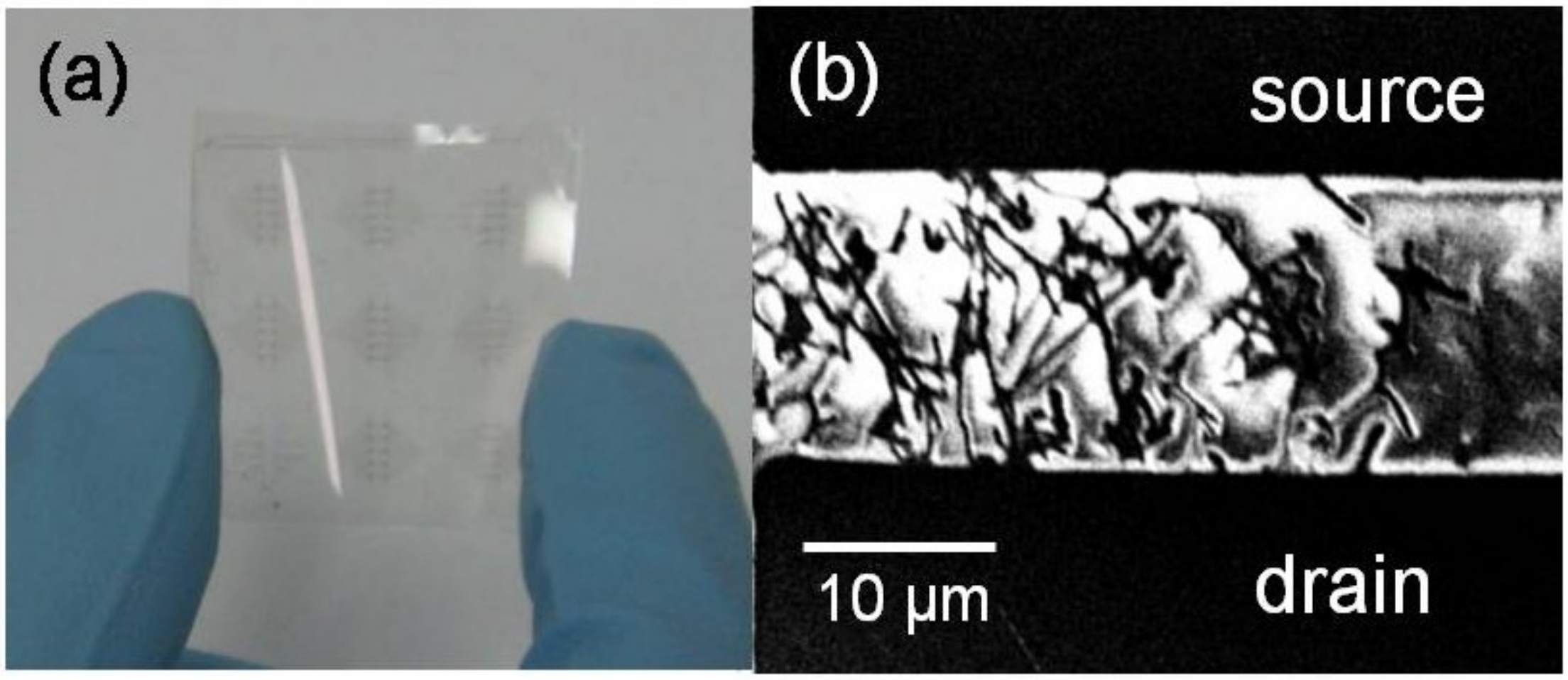

Figure 2. a) An optical image of a one inch square PET substrate with 90 CNT electrode sub-assemblies. b) A SEM image of the device channel of an all-CNT TFT. Airbrushed CNT source-drain electrodes are spanned by a sparse network of CVD grown CNTs. CNTs show poor contrast due to charging of the plastic substrate.

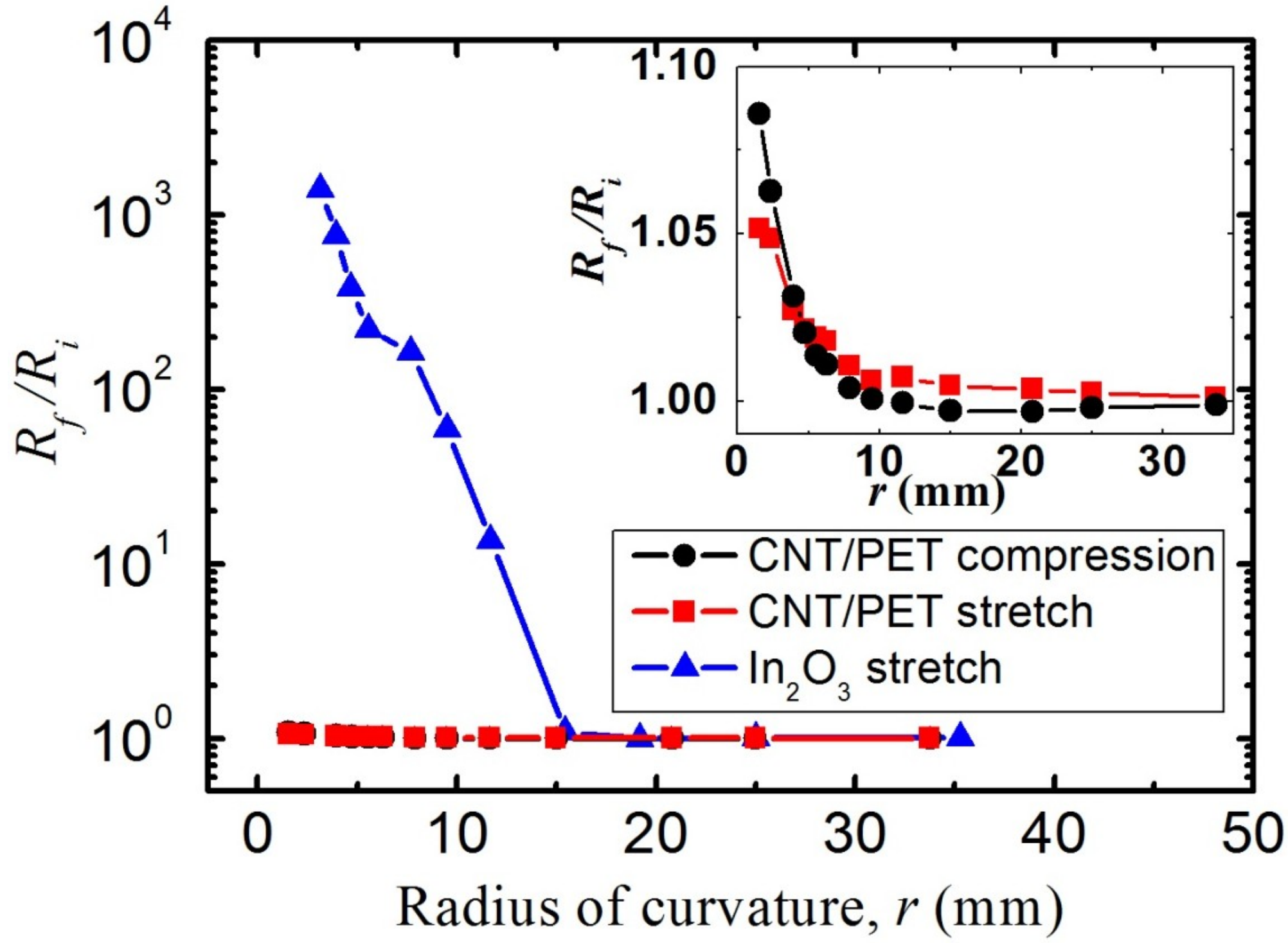

Figure 3. The ratio of the sheet resistances, $R_f/R_i$, before ($R_i$) and after bending ($R_f$) as a function of the radius of curvature ($r$) for airbrushed CNT thin-film as well as $In_2O_3$ on PET substrate. Black circles indicate compressive stress, red squares indicate tensile stress. Bending the $In_2O_3$ thin-film results in three orders of magnitude increase in sheet resistance at $r$ = 3 mm, while the sheet resistance of CNT thin-films increases only by 7 % at $r$ = 2 mm (see inset).

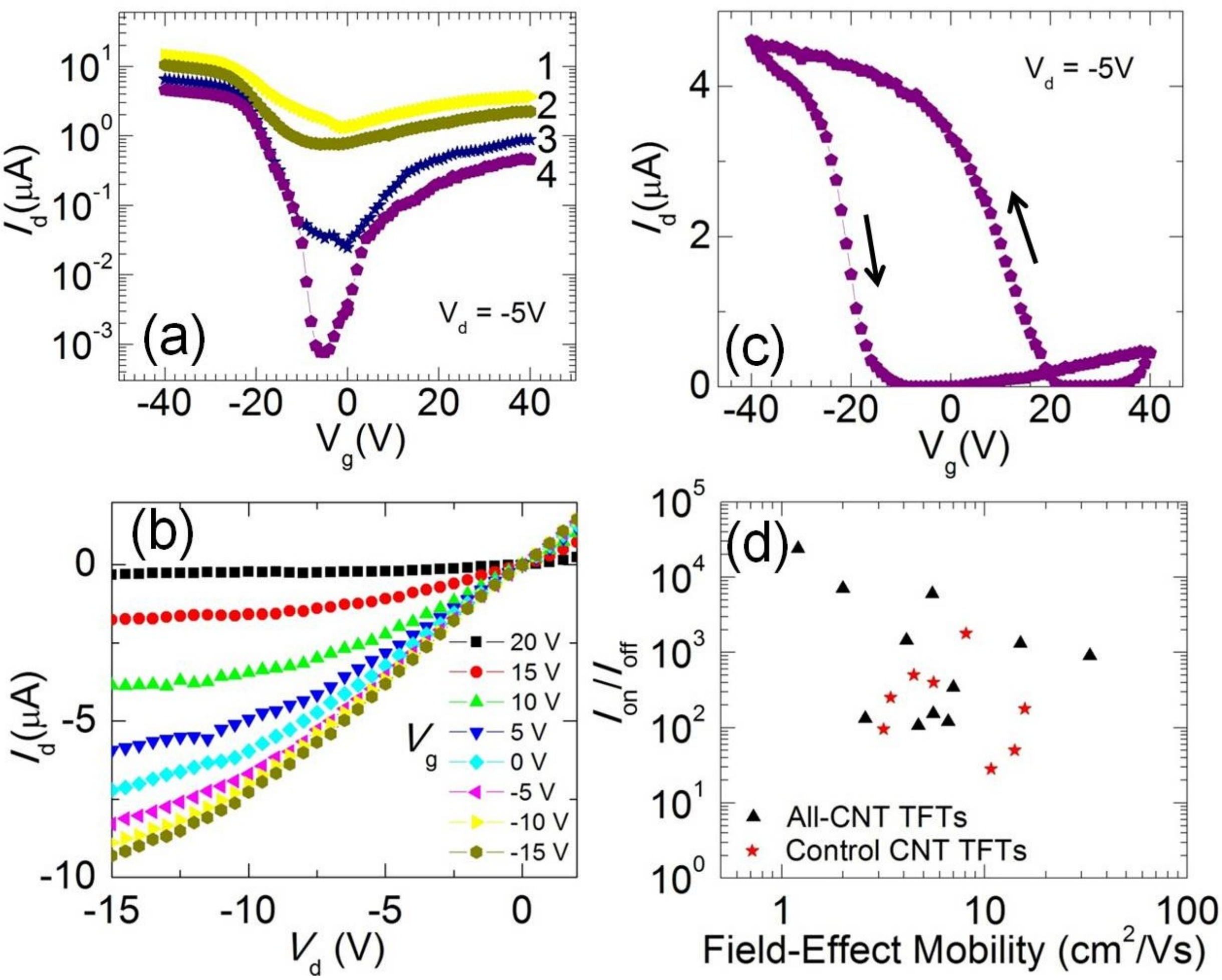

Figure 4. a) Transfer characteristics of an all-CNT device ($L$ = 60 μm) device at $V_d$ = -5 V at intermediate stages of electrical breakdown as the gate bias is sweeped from -40 V to 40 V. The on/off ratio was increased by 3 orders of magnitude by successively burning metallic CNTs at $V_g$ = 30 V (curves 1 through 4). (b) Output characteristics and (c) transfer characteristics of the device after electrical burning. (d) On/off ratio as a function of field-effect mobility for CNT-based devices as well as control CNT devices with $L$ varying in range 10 – 100 μm.